\documentstyle[art10,epsf,makeidx,bezier]{article}
\def\si{\quad}
\def\sii{\qquad}

\def\siiii{\qquad\qquad}
\def\b:{\begin{equation}}
\def\b:{\begin{equation}}
\def\b:{\begin{equation}}
\def\e:{\end{equation}}
\def\be:{\begin{eqnarray}}
\def\ee:{\end{eqnarray}}
\def\ba{\begin{array}}
\def\ea{\end{array}}
\def\nn{\nonumber\\}

\def\bmin{\begin{minipage}{15cm} }
\def\emin{\end{minipage}}
\newcommand{\lb}[1]{\label{eqn:#1}}
\newcommand{\rf}[1]{\ref{eqn:#1}}
\def\and:{&&\!\!\!}

\def\1:{f_{1}}
\def\2:{f_{2}}
\def\3:{f_{3}}
\def\4:{f_{4}}

\begin{document}
\baselineskip 20pt
\vglue 2cm
\centerline{\Large {\bf Symmetric Linear B\"acklund Transformation}}
\centerline{\Large {\bf for Discrete BKP and DKP Equations}}

\vglue 4cm
\centerline{Nobuhiko SHINZAWA}
\vglue 1cm
\centerline{{\it Department of Physics, Tokyo Metropolitan University}}
\centerline{{\it Minamiohsawa 1-1, Hachiohji, Tokyo 192-03, Japan}}
\centerline{{\it email: nobuhiko@phys.metro-u.ac.jp}}
\vglue 6cm
\begin{abstract}

Proper lattices for the discrete BKP and the discrete DKP equaitons are
determined.
Linear B\"acklund transformation equations
for the discrete BKP and the DKP equations 
are constructed,
which possesses
the lattice symmetries and generate
auto-B\"acklund transformations

\end{abstract}

\vfill\eject

\section{Introduction}
Discrete counterpart of the KP equation, known in the name of discrete KP
equation or Hirota Bilinear Difference Equation (HBDE)\cite{Miwa}\cite{KP}\cite{Hirota}, plays a central role
in the study of integrable nonlinear systems. It embodies infinite number of
integrable differential equations. It is satisfied by string correlation
function in particle physics\cite{SS}, and is also satisfied by transfer matrices of
some solvable lattice model \cite{KLWZ}.

The equation possesses auto-B\"acklund transformation which acts on the
solutions to the discrete KP equation and turns n-soliton solution to
n+1-soliton solution \cite{Hirota}. This transformation is generated by a pair of linear
equations.I call the equations Linear B\"acklund Transformation Equation
(LBTE) in this paper.
They also play the important role in the study of integrable systems.
They are Lax pair of the discrete KP equation.
They generate Bethe ansatz solution of some solvable lattice model.
\cite{SS}

On the other hand, from the structure of the discrete KP equation,
dependent variables of discrete KP equation are considered to reside on face
centered cubic (FCC) lattice.
Indeed, the discrete KP equation does not change its form
under the rotation of the FCC lattice.
However, LBTE to the discrete KP equation changes those form under the
rotation.

In the previous paper \cite{NSSS},
the LBTE are extended to possess the lattice symmetry,
which I call the Symmetric LBTE in this paper
and the following results are obtained.
The extended equations were arranged in the form of\\
matrix $\times$ vector$=0$\\
by considering the lattice symmetry.
The condition that the extended equations
have nontrivial solutions, i.e. vanishment of the determinant of the matrix,
is just the discrete KP equation.
Furthermore the extended equations were generalized to higher dimension due
to the lattice symmetry itself.

On the one hand,
comparing with the KP hierarchy which possesses $A_{\infty}$ type
Lie group symmetry acting on the space of solutions \cite{KP},
there are integrable hierarchies
which possess the $B_{\infty}$ or $D_{\infty}$ Lie group symmetry
behind them \cite{Kac} \cite{BKP}.
Such hierarchies are known in the name of BKP,
fermionic BKP and DKP hierarchy.
To these hierarchies, one can find discrete equations
corresponding to the discrete KP equation by considering the
infinite dimensional symmetry behind them.
What would happen, when one corresponds proper lattices to the discrete
equations, and consider the lattice symmetry of LBTE's to these discrete
equations?

In this paper, I show the following results obtained by considering above
question.

First, proper lattices are determined by comparing the type of the
infinite dimensional symmetries and the dimension of the discrete
equations. The discrete equations are invariant under the rotation of the
lattices.

Second, the LBTE to the discrete equation which possesses the lattice
symmetry are constructed.
The LBTE can be represented by the functions on the
vertices of some regular polyhedrons in the lattice.

Third,
just the discrete equation is derived from the consistency
condition for LBTE,
when one considers following two types of consistency condition.
One condition is determinant type consistency condition explained
above and another one is condition
similar to the compatibility condition for Lax pair,
which is called compativility type condition in this paper.
In some cases,
LBTE can not be arranged in the form of matrix $\times$ vector=0, however,
in
such cases, compatibility type condition becomes just the discrete equation.
Furthermore, by considering these two types of conditions, one can
check that the LBTE generates auto-B\"acklund transformation for the discrete
equation.

After explaining main idea in the case of the KP hierarchy in section 2,
 I show the above
results in the case of the BKP, fermionic BKP and DKP hierarchy
 in section 3 and 4.

\section{Case of the KP hierarchy}
\unitlength 1mm
\begin{picture}(120,45)
\put(-5,0){\epsfxsize=11cm\epsfbox{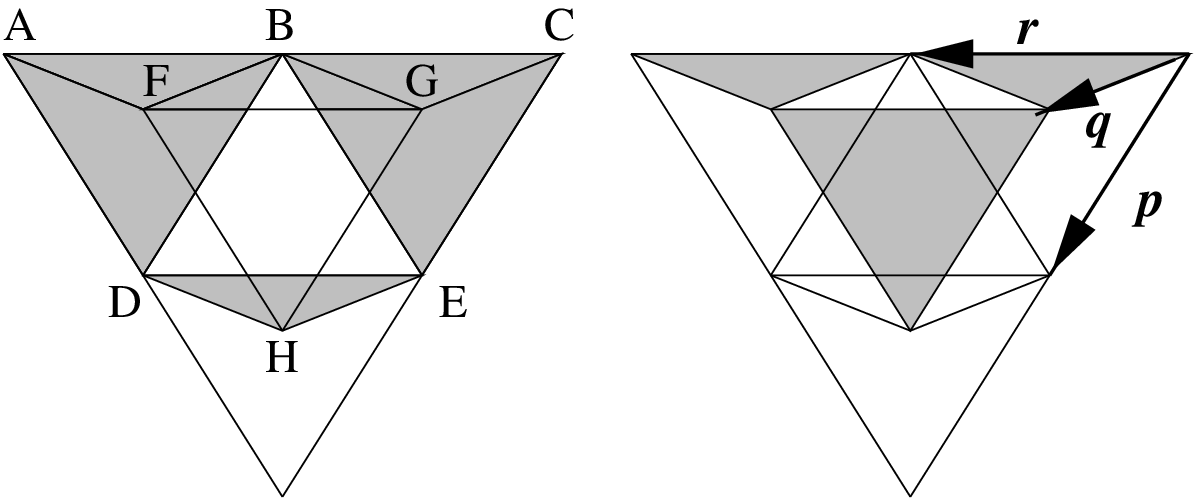}}
\end{picture}

$A_{\infty}$ symmetry of space of solutions to the KP hierarchy allows one
to deal it in simple form \cite{Sato}.
This section is devoted to summarize some results obtained in the previous
paper \cite{NSSS} , concerning to the discrete formula for the KP hierarchy
and its LBTE.

Lowest dimensional discrete formula belonging to the KP hierarchy,
known in the name of discrete KP equation,
is a 3 dimensional equation \cite{Hirota}\cite{Miwa}\cite{KP}.
(See also Appendix)
The equation is most fundamental equation among discrete formulae for the KP
hierarchy,
in the sense that higher dimensional discrete formulae can be decomposed to
it.
Discrete KP equation is the following.
\be:
z_{12}z_{34}f(k_{1}+1,k_{2}+1,k_{3},k_{4})f(k_{1},k_{2},k_{3}+1,k_{4}+1)\nn
-z_{13}z_{24}f(k_{1}+1,k_{2},k_{3}+1,k_{4})f(k_{1},k_{2}+1,k_{3},k_{4}+1)\nn
+z_{14}z_{23}f(k_{1}+1,k_{2},k_{3},k_{4}+1)f(k_{1},k_{2}+1,k_{3}+1,k_{4})
=0
\lb{dKP}
\ee:
Here, $z_{ij}$ is
\be:
z_{ij}=z_{i}-z_{j}
\ee:
and $z_{i}(i=1,2,3,4)$ are arbitrary complex constants.
The equation seems defined on 4 dimensional lattice space.
However, since the sum of the four variables does not vary in the equation,
the equation is actually three dimensional equation.
Indeed, the following variable transformation allows one to represent the
equation by three variables,
\be:
p=k_{1},\si q=k_{2},\si r=k_{3},\si n=k_{1}+k_{2}+k_{3}+k_{4}
\lb{vari}
\ee:
.

First, I seek the proper lattice for the discrete KP equation.
To find it, I relate a regular polyhedron to the discrete KP equation,
and join the polyhedrons.
The proper polyhedron is considered to be octahedron,
because the equation connects f's on six points.
In fact, when one considers the four variables to be orthogonal,
the equation takes the form of a summation of the products of f on one
vertex of octahedron and another f on opposite side vertex.
By joining octahedrons, a FCC lattice is constructed.
Thus, proper lattice for the discrete KP equation turns out to be FCC
lattice.

It is known that there exist equations which generate B\"acklund
transformation \cite{Hirota}\cite{SS}.
In the previous paper \cite{NSSS}, the equations are extended to possess the
lattice symmetry of FCC lattice.
Each of the equations takes same form with discrete formula of modified KP
hierarchy \cite{KP}.(See Appendix)
The equations take the following form.
\be:
\sum_{j,k,l=1}^{4}\frac{1}{2}\epsilon_{ijkl}z_{kl}f_{kl} g_{j}=0,
\si  i=1,2,3,4 \si
\ee:
Where, $\epsilon_{ijkl}$ is the Levi-Chivita tensor and small indices on f and g mean to increase the variable corresponding
to index by one.

For example,
\be:
g_{1}=g(k_{1}+1,k_{2},k_{3},k_{4})
\si f_{12}=f(k_{1}+1,k_{2}+1,k_{3},k_{4})
\si f_{11}=f(k_{1}+2,k_{2},k_{3},k_{4})
\ee:
I called these equations Symmetric LBTE, in the previous paper.

The Symmetric LBTE can be related to fundamental regular polyhedrons in FCC lattice,
when one arranges these equations suitably.
Fundamental regular polyhedrons in FCC lattice are octahedron and two
tetrahedrons.
There are two such arrangements.
One of which is as follows.
\be:
\left(
\begin{array}{cccc}
0&z_{34}f_{34}&-z_{24}f_{24}&z_{23}f_{23}\\
-z_{34}f_{34}&0&z_{14}f_{14}&-z_{13}f_{13}\\
z_{24}f_{23}&-z_{14}f_{14}&0&z_{12}f_{12}\\
-z_{23}f_{23}&z_{13}f_{13}&-z_{12}f_{12}&0\\
\end{array}\right)
\left(
\begin{array}{cccc}
g_{1}\\
g_{2}\\
g_{3}\\
g_{4}
\end{array}\right)
=0
\lb{kplb}
\ee:
Another arrangement is to construct the elements of matrix and vector with g
and f.
In both of the arrangements, f and g in the matrix elements are on vertices
of an octahedron.
On the other hand, each g and f in the vector elements are on vertices of a
tetrahedron dual to each other.
Note that the arrangements make the lattice symmetry of the equations
manifest.

Here is a further remark.
FCC lattice is a root lattice of $A_{3}$ type Lie group, and two
tetrahedrons
and
one octahedron is weight diagrams of three fundamental representations of
$A_{3}$ type Lie
group. This implies that one can determine  the regular polyhedrons and
proper lattice
for the discrete formula, by comparing the dimension of the equation and
symmetric groups corresponding to the hierarchy.

In what follows, I show that Symmetric LBTE generate B\"acklund
transformation successfully.
Here, ``successfully'' means the following:
Whenever f satisfies the discrete KP equation,
 the Symmetric LBTE can be solved for g, and
the solution solves the discrete KP equation automatically.

First, both f and g in the Symmetric LBTE satisfy discrete KP equation when
they satisfy Symmetric LBTE.
To show it, we consider equation $(\rf{kplb})$ as 4 linear equations to be
solved for four g's.
This four linear equations can be solved only if the Determinant of the
coefficient matrix vanishes.
In this case, one can use the fact that Determinant of antisymmetric matrix
is a square of Pfaffian.
This fact leads the condition into the following compact form.
\be:
Det(a_{ij})=(Pfaff(a_{ij}))^{2}\nn
=(z_{12}z_{34}f_{12}f_{34}
-z_{13}z_{24}f_{13}f_{24}
+z_{14}z_{23}f_{14}f_{23})^{2}
=0
\ee:
This is nothing but the discrete KP equation.
Now, one can exchange the role of f and g, by translating each equation in
($\rf{kplb}$),
so that g also satisfies the discrete KP equation.
Remark, that this explanation depends essencially on the arrangement of the
Symmetric LBTE.

Second, Symmetric LBTE can be solved for g whenever f satisfies the discrete
KP equation.
This fact and first statement show that Symmetric LBTE generate B\"acklund
transformation successfully.
However,it requires little more consideration.
In fact, other consistency condition than determinant type consistency
condition arises as follows.
Consider the determinant type consistency condition is already satisfied.
Then, since the rank of the coefficient matrix is 2,
two equations out of four equations in $(\rf{kplb})$ remain independent.
I chose such two equations as first two equations in ($\rf{kplb}$).
For simplicity, I express the equations by the variables $p,q,r$ in
($\rf{vari}$).
\be:
1:\si \and:
z_{34}f_{r}g_{q}-z_{24}f_{q}g_{r}+z_{23}f_{qr}g=0\nn
2: \si \and:
-z_{14}f_{p}g_{r}+z_{13}f_{pr}g+z_{34}f_{r}g_{p}=0
{\lb{ink}}
\ee:
These two equations provide several ways to obtain g on one point from g on
other points.
The value of obtained g needs to be independent on the choice of the ways.
For example, in figure 1, one can require the value of g(H) from g(A), g(B),
g(C), according to following two procedures.
\be:
1 \si
g(A),g(B) \stackrel{1}{\rightarrow} g(F)\siiii \siiii \siiii \nn
\si \si g(B),g(C) \stackrel{1}{\rightarrow} g(G) \si
\stackrel{2}{\rightarrow} \si g(F),g(G)\rightarrow g(H)\nn
2 \si
g(A),g(B)\stackrel{2}{\rightarrow} g(D)\siiii \siiii \siiii \nn
\si \si g(B),g(C)\stackrel{2}{\rightarrow} g(F) \si
\stackrel{1}{\rightarrow} \si g(D),g(F)\stackrel{1}{\rightarrow} g(H)\ee:
One can evaluate the value of g on one point from the values of g on two
points
by using the equation corresponding to the number over the arrow
($\rf{ink}$).Here, arrow represents that process.
After some calculation, two procedures lead the following expressions of
g(H).
\be:
1: \and: g(H)=g(A)\frac{z_{14}z_{24}f_{pq}f_{qr}}{z_{34}z_{34}f_{qr}f_{rr}}
+g(C)\frac{z_{31}z_{32}f_{pqr}f_{qr}}{z_{34}z_{34}f_{qr}f_{rr}}
+g(B)(\frac{z_{14}z_{32}f_{pq}f_{qrr}}{z_{34}z_{34}f_{qr}f_{rr}}
+\frac{z_{31}z_{24}f_{pqr}f_{q}}{z_{34}z_{34}f_{qr}f_{r}})\nn
2: \and:
g(H)=g(A)\frac{z_{14}z_{24}f_{pq}f_{qr}}{z_{34}z_{34}f_{qr}f_{rr}}
+g(C)\frac{z_{31}z_{32}f_{pqr}f_{qr}}{z_{34}z_{34}f_{qr}f_{rr}}
+g(B)(\frac{z_{24}z_{32}f_{pq}f_{prr}}{z_{34}z_{34}f_{pr}f_{rr}}
+\frac{z_{32}z_{14}f_{pqr}f_{p}}{z_{34}z_{34}f_{pr}f_{r}})
\lb{two}
\ee:
These two g(H)'s need to coincide.
By equating the coefficients of $g(B)$, one can express the condition as
follows.
\be:
\frac{f_{pq}f_{pqr}}{z_{3}^{2}f_{qr}f_{pr}} \times
(e^{\partial_{r}}-1)(\frac{Discrete \si KP}{f_{r}f_{pq}})=0
\ee:
Here, $e^{\partial_{i}}$ acts on arbitrary functions from leftside and
increases the variable corresponding to index $i$ by one in the function.
In the equation $Discrete \si KP$ expresses the right hand side of equation
(${\rf{dKP}}$).
The coefficients of $g(A),g(C)$ automatically coincide.
Thus, in this case, the condition is satisfied if f satisfies the discrete
KP equation.

However, this compatibility condition guaranties the existence of solution
of g to Symmetric LBTE.
To show it, I first put initial value of g on the plane p+q=c.
Here c is arbitrary constant.
Using the two equations in $(\rf{two})$ independently,
one can obtain g's on the plane p+q=c+1.
To make the value of g on the plane p+q=c+1 unique,
I impose the following condition on initial value of g.
\be:
\frac{-z_{24}f_{pq}g_{pr}
+z_{23}f_{pqr}g_{p}}
{z_{3}f_{pr}}
=
\frac{-z_{14}f_{pq}g_{qr}+z_{13}f_{pqr}g_{q}}
{z_{34}f_{qr}}
\lb{jyouken}
\ee:
If the compatibility condition is satisfied, this condition is also
satisfied by g's on the plane p+q=c+1.
Therefore one can inductively construct g's on all lattice point, if the
compatibility condition is satisfied.
This concludes that there exist at least one solution of g for Symmetric
LBTE
with f being a solution to the discrete KP equation.

\section{Case of the BKP and the fermionic BKP hierarchy}
\unitlength 1mm
\begin{picture}(50,50)
\put(-5,0){\epsfxsize=5cm\epsfbox{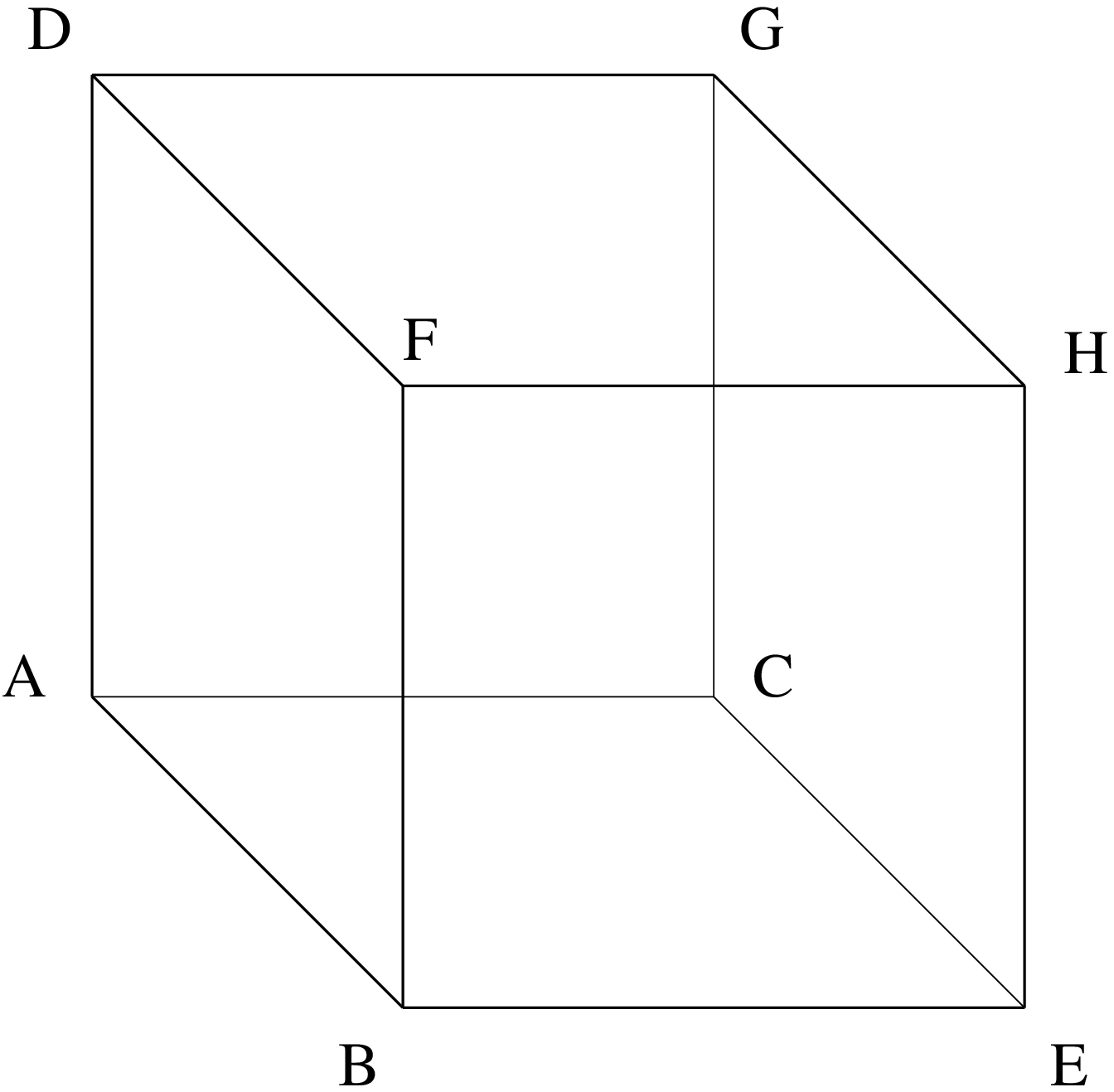}}
\end{picture}

Lowest dimension discrete formulae belonging to the BKP and the fermionic BKP
hierarchy are known to be as follows \cite{Miwa}\cite{BKP}\cite{Kac}.
(See Appendix)
\be:
a_{12}b_{23}a_{31}f_{1}f_{23}
+a_{12}a_{23}b_{31}f_{2}f_{13}
+b_{12}a_{23}a_{31}f_{3}f_{12}
+b_{12}b_{23}b_{31}f f_{123}=0
\lb{dbkp}
\ee:
Here, $a_{ij}$ and $b_{ij}$ are constant coefficients, and can be expressed
as
\be:
BKP&&
a_{ij}=z_{i}+z_{j},\si
b_{ij}=z_{i}-z_{j}\nn
fermionic \si BKP&&
a_{ij}=1,\si
b_{ij}=z_{i}-z_{j}
\lb{coe}
\ee:
respectively.
Here, $z_{i}(i=1,2,3)$ are arbitrary complex constants.
The equation is called discrete BKP equation, for the first choice of
coefficient in ($\rf{coe}$).
However I call the equation discrete BKP equation for both cases, neglecting
the difference of the coefficients.

By considering the modified BKP hierarchy and the modified fermionic BKP
hierarchy \cite{BKP} \cite{Kac}, one can obtain LBTE for the equations.
Collecting the discrete formulae which depend on variables appearing in the
discrete BKP equation, in the modified BKP hierarchy, the following three
equations can be obtained.(See Appendix)
\be:
a_{ij}\bigl( f_{j}g_{i}-f_{i}g_{j}\bigr)=
b_{ij}( f g_{ij}-f_{ij}g\bigr) \nn
i,j=1,2,3 \si i\neq j
\ee:

The discrete BKP equation is a three dimensional equation and connect f's on
eight points.
Thus the cube may become proper regular polyhedron for discrete BKP
equation.
Indeed, when the three variables p,q,r are considered to be orthogonal,
the equation takes the form of summation of products of f on one vertex of
the cube and another f on opposite side vertex.
What is obtained by joining cubes is a simple cubic lattice.
Therefore I argue that the simple cubic lattice is the proper lattice for
the discrete BKP equation.
Note that a cube and simple cubic lattice are obtained as weight diagram of
a
fundamental representation and root lattice for $B_{3}$-type Lie group,
respectively.

On the other hand each of the equations in LBTE connects f's and g's on a
square. It does not change its form under the exchange of f and g.

As in the case of the discrete KP equation,
discrete BKP equation appears as consistency condition for the
LBTE, by arranging the LBTE in symmetric form.
In simple cubic lattice, fundamental regular polyhedron is a cube.
Thus such symmetric equations may connect f and g on a cube.
Indeed, by collecting equations in LBTE corresponding to 6 squares of a cube,
one can make such equations.
These equations do not take the form of matrix times vector.
Thus determinant type consistency condition does not arise in this case.
However compatibility condition may arise because one cube contains two
parallel squares.
Indeed, there are three ways to obtain $g_{123}$ from $g,g_{1},g_{2},g_{3}$
by using the equations.
\be:
1:\and: g,g_{1},g_{2} \rightarrow g_{12} \siiii \siiii \siiii \si \nn
\and:  g,g_{2},g_{3} \rightarrow g_{23} \sii g_{2},g_{12},g_{23} \rightarrow
g_{123}\nn
2:\and: g,g_{2},g_{3} \rightarrow g_{23}\siiii \siiii \siiii \si\nn
\and: g,g_{1},g_{3} \rightarrow g_{13} \sii g_{3},g_{23},g_{13} \rightarrow
g_{123}\nn
3:\and: g,g_{1},g_{2} \rightarrow g_{12}\siiii \siiii \siiii \si\nn
\and: g,g_{1},g_{3} \rightarrow g_{13} \sii g_{1},g_{12},g_{13}\rightarrow
g_{123}\nn
\ee:
After some calculation these procedures lead to the following expressions of
$g_{123}$.
\be:
\!\!\! \and: 1: \,
g_{123}=g\times 0
+g_{1}\bigr( \frac{a_{31}a_{12}f_{23}}{-b_{31}b_{12}f}\bigr)
+ g_{2}\bigr( \frac{a_{31}a_{12}f_{23}f_{1}}{b_{31}b_{12}f_{2}f}
-\frac{a_{13}a_{23}f_{12}f_{3}}{b_{13}b_{23}f_{2}f}+\frac{f_{123}}{f_{2}}
\bigr)
-g_{3}\bigl( \frac{a_{13}a_{23}f_{12}}{b_{31}b_{23}f} \bigr)\nn
\!\!\! \and: 2: \,
g_{123}=g \times 0
-g_{1}\bigr(\frac{a_{31}a_{12}f_{23}}{b_{31}b_{12}f}\bigr)
-g_{2} \bigl( \frac{a_{12}a_{23}f_{13}}{b_{12}b_{23}f} \bigl)
+g_{3}\bigl( \frac{a_{12}a_{23}f_{13}f_{2}}{b_{12}a_{23}f_{3}f}
+\frac{a_{12}a_{31}f_{23}f_{1}}{b_{12}b_{31}f_{3}f}+\frac{f_{123}}{f_{3}}
\bigr)
\ee:
These three expressions need to coincide.
The condition becomes just the discrete BKP equation for f, after equating
the coefficients of $g,g_{1},g_{2},g_{3}$.

In this case, each equation in the LBTE does not change it's form under the
exchange of f and g.
This means that g also needs to satisfy discrete BKP equation, to be able to
solve the LBTE for f.
Therefore the LBTE can be solved for g only when f satisfies discrete BKP
equation, and the solution also satisfies discrete BKP equation
automatically.

To verify the existence of solution of g to LBTE, I construct a surface on
which initial value of g is put and move the surface, as before.
From the structure of the LBTE, such surface turns out to be constructed
from two plane $k_{1}+k_{2}+k_{3}=c$ and $k_{1}+k_{2}+k_{3}=c+1$.
Here c is arbitrary integer. 
By using three equations in LBTE independently, one can obtain the value of
g on $k_{1}+k_{2}+k_{3}=c+2$.
Thus, the conditions to make these procedure equivalent need to be imposed
on initial value of g on two planes $k_{1}+k_{2}+k_{3}=c$ and
$k_{1}+k_{2}+k_{3}=c+1$.
\be:
&&\frac{f_{123}}{f_{3}}g_{3}
+\frac{a_{12}}{b_{12}f_{3}}\bigl( b_{12}f_{12}g_{13}-f_{13}g_{12}\bigr)\nn
&=&\!\!\!\frac{f_{123}}{f_{1}}g_{1}
+\frac{a_{23}}{b_{23}f_{1}}\bigr( b_{23}f_{13}g_{12}-f_{12}g_{13}\bigr)\nn
&=&\!\!\!\frac{f_{123}}{f_{2}}g_{2}
+\frac{a_{31}}{b_{31}f_{2}}\bigl( b_{23}f_{12}g_{23}-f_{23}g_{12}\bigr)\nn
\lb{bscon}
\ee:
These conditions do require no consistency condition.
Namely one can find initial value that satisfy these conditions.
Such initial value of g provides g on $k_{1}+k_{2}+k_{3}=c+2$.
Thus,
by showing g on $k_{1}+k_{2}+k_{3}=c+1$ and $k_{1}+k_{2}+k_{3}=c+2$
automatically satisfies the condition ($\rf{bscon}$),
one can verify the existence of solution of g to LBTE recursively.
That condition becomes just the compatibility condition.
As a consequence, one can verify that there is at least one solution of g
for LBTE whenever f satisfies the discrete BKP equation.

\section{Case of the DKP hierarchy}

Lowest dimensional discrete formulae which belong to the DKP hierarchy
\cite{Kac} are the following two equations.
\be:
z_{14}z_{23}f_{23}f_{14}
-z_{13}z_{24}f_{13}f_{24}
+z_{12}z_{34}f_{12}f_{34}
-z_{12}z_{13}z_{14}z_{23}z_{24}z_{34}f_{1234}f=0\nn
z_{23}z_{24}z_{34}f_{234}f_{1}
-z_{13}z_{14}z_{24}f_{134}f_{2}
+z_{12}z_{14}z_{24}f_{124}f_{3}
-z_{12}z_{13}z_{23}f_{123}f_{4}=0
\ee:
where $z_{ijk}=z_{ij}z_{jk}z_{ki}$ .
In these equations, f represents $\tau$-function for DKP hierarchy,
but some variable transformation is performed. (See Appendix)
We call these equations discrete DKP equations.

I first seek the regular polyhedrons and proper lattice for these equations.
In this case, one can not see figures, since the equations are defined on
four dimensional lattice space.
However, consideration in the previous cases suggests that one can find it
by considering representations of $D_{4}$-type Lie group.
Namely, the regular polyhedrons may be obtained as weight diagrams of
fundamental representations for $D_{4}$-type Lie group, and the proper
lattice may be the root lattice for the group.
The Dynkin diagram for $D_{4}$-type Lie group is as follows.

\unitlength 1mm
\begin{picture}(50,50)
\put(-5,0){\epsfxsize=5cm\epsfbox{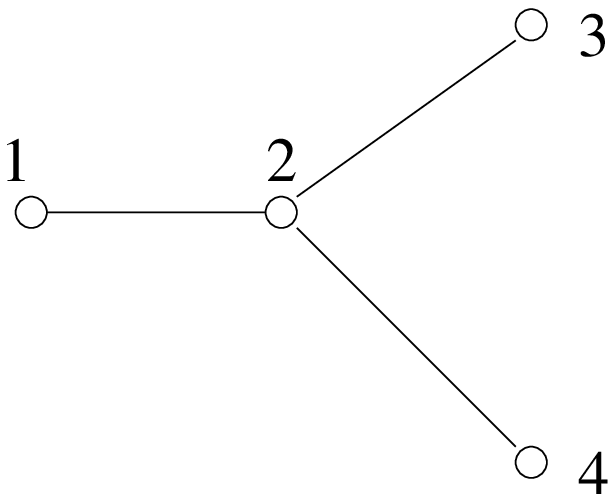}}
\end{picture}

\noindent
The root vectors corresponding to the indices in the figure can be
represented in Cartesian coordinate as follows.
\be:
{\bf \alpha}_{1}=(1,-1,0,0),{\bf\alpha}_{2}=(0,1,-1,0),
{\bf\alpha}_{3}=(0,0,1.-1),{\bf\alpha}_{4}=(0,0,1,1)
\ee:
In this choice of coordinate, weight diagrams of three fundamental
representations corresponding to Dynkin indices
\\$[1,0,0,0],[0,0,1,0],[0,0,0,1]$\\  become the following three 4
dimensional reguler 16-hedrons.
\be:
[1,0,0,0] \si \;\;\; [0,0,1,0] \si [0,0,0,1]\nn
(1,0,0,0) \si \;\; (1,1,1,0) \si (1,1,1,1)\nn
(0,1,0,0) \si \;\; (1,1,0,1) \si (1,1,0,0)\nn
(0,0,1,0) \si \;\; (1,0,1,1) \si (1,0,1,0)\nn
(0,0,0,1) \si \;\; (0,1,1,1) \si (1,0,0,1)\nn
(-1,0,0,0) \si (1,0,0,0) \si (0,1,1,0)\nn
(0,-1,0,0) \si (0,1,0,0) \si (0,1,0,1)\nn
(0,0,-1,0) \si (0,0,1,0) \si (0,0,1,1)\nn
(0,0,0,-1) \si (0,0,0,1) \si (0,0,0,0)\nn
\lb{16}
\ee:
Two Discrete DKP equations connect f's on [0,0,1,0] type and [0,0,0,1] type
regular 16-hedrons respectively.
Therefore, proper lattice for discrete DKP equations turns out to be the
root lattice for $D_{4}$-type Lie group.

Collecting the discrete formulae that depend on variables arising in
discrete DKP equations, in modified DKP hierarchy \cite{Kac},  one can
obtain following 8 linear equations.
\be:
\sum_{jkl}(\frac{1}{2}\epsilon_{ijkl}z_{kl}f_{kl}g_{j}+\frac{1}{6}z_{jkl}fg_
{jkl})=0\nn
\sum_{jkl}(\frac{1}{2}\epsilon_{ijkl}z_{jk}f_{l}g_{jk}+\frac{1}{6}z_{jkl}f_{
jkl}g)=0
\lb{dlbte}
\ee:
In each of the equations in ($\rf{dlbte}$), $f$ and $g$ are difined on two
tetrahedrons dual to each other.
8 tetrahedrons arise in both equations  of  ($\rf{dlbte}$).
We study how these tetraherons appear in root lattice for $D_{4}$-type Lie
group, to arrange these equations in symmetric form.
There are 24 tetrahedrons in the root lattice for $D_{4}$-type Lie group.
Three regular 16-hedrons in ($\rf{16}$) contain 16 tetrahedrons in it.
8 tetrahedrons arising in ($\rf{dlbte}$) are the tetrahedrons which are
simultaneously contained in [0,0,1,0] and [0,0,0,1] tetrahedrons.
Therefore, by arranging the equations to the form in which g is on [0,0,1,0]
or [0,0,0,1] regular 16-hedron, the symmetric arrangements may be obtained.
In fact, in each of the case, f turns out to be on [0,0,0,1] and [0,0,1,0]
regular 16 hedrons respectively.
Moreover the equations can be summarized in the form of regular matrix times
vector.
\be:
\left[
\begin{array}{cccccccc}
0&z_{34}f_{34}&z_{42}f_{24}&z_{23}f_{23}&z_{234}f&0&0&0\\
z_{34}f_{34}&0&z_{41}f_{14}&z_{13}f_{13}&0&z_{341}f&0&0\\
z_{24}f_{24}&z_{41}f_{14}&0&z_{12}f_{12}&0&0&z_{412}f&0\\
z_{23}f_{23}&z_{31}f_{13}&z_{12}f_{12}&0&0&0&0&z_{123}f\\
z_{234}f_{1234}&0&0&0&0&z_{34}f_{34}&z_{42}f_{24}&z_{23}f_{23}\\
0&z_{341}f_{1234}&0&0&z_{34}f_{34}&0&z_{41}f_{14}&z_{13}f_{13}\\
0&0&z_{412}f_{1234}&0&z_{24}f_{24}&z_{41}f_{14}&0&z_{12}f_{12}\\
0&0&0&z_{123}f_{1234}&z_{23}f_{23}&z_{31}f_{13}&z_{12}f_{12}&0
\end{array}\right]
\times
\left[
\begin{array}{cccccccc}
g_{1}\\
g_{2}\\
g_{3}\\
g_{4}\\
g_{234}\\
g_{134}\\
g_{124}\\
g_{123}\\
\end{array}
\right]
=0
\lb{lbte1}
\ee:
\be:
\left[
\begin{array}{cccccccc}
0&z_{13}f_{4}&z_{41}f_{3}&z_{341}f_{134}&z_{34}f_{1}&0&0&0\\
z_{12}f_{4}&0&z_{41}f_{2}&z_{412}f_{124}&0&z_{24}f_{1}&0&0\\
z_{12}f_{3}&z_{31}f_{2}&0&z_{123}f_{123}&0&0&z_{23}f_{1}&0\\
z_{34}f_{134}&z_{42}f_{124}&z_{23}f_{123}&0&0&0&0&z_{234}f_{1}\\
z_{34}f_{234}&0&0&0&0&z_{13}f_{123}&z_{41}f_{124}&z_{341}f_{2}\\
0&z_{24}f_{234}&0&0&z_{12}f_{123}&0&z_{41}f_{134}&z_{412}f_{3}\\
0&0&z_{23}f_{234}&0&z_{12}f_{124}&z_{31}f_{134}&0&z_{123}f_{4}\\
0&0&0&z_{234}f_{234}&z_{34}f_{2}&z_{42}f_{3}&z_{23}f_{4}&0
\end{array}
\right]
\times
\left[
\begin{array}{cccccccc}
g_{12}\\
g_{13}\\
g_{14}\\
g\\
g_{34}\\
g_{24}\\
g_{23}\\
g_{1234}\\
\end{array}
\right]
=0\nn
\lb{lbte2}
\ee:

These arrangements make the lattice symmetry of the equations manifest.

In this case, the symmetric arrangements induce two determinant type
consistency conditions, since they take the form of matrix times vector.
One finds that these two determinants are just the four square of two
discrete DKP equations, after some calculation.
\be:
\bigr(z_{14}z_{23}f_{23}f_{14}
-z_{13}z_{24}f_{13}f_{24}
+z_{12}z_{34}f_{12}f_{34}
-z_{12}z_{13}z_{14}z_{23}z_{24}z_{34}f_{1234}f \bigl)^{4}=0\nn
\bigr(z_{23}z_{24}z_{34}f_{234}f_{1}
-z_{13}z_{14}z_{24}f_{134}f_{2}
+z_{12}z_{14}z_{24}f_{124}f_{3}
-z_{12}z_{13}z_{23}f_{123}f_{4} \bigl)^{4}=0
\ee:
Thus, two discrete DKP equations arise as consistency conditions for LBTE.
This concludes that, when one solves the equations for g provided f being a
solution for discrete DKP equation, g also satisfies discrete DKP equations,
since the equations are invariant under the exchange of f and g.

One can prove the existence of solution for g to LBTE with f being a
solution to discrete DKP equations,
by using the same method as in the previous cases.
Namely, one can consistently evaluate g from initial value of g defined on
some surface, whenever f satisfies discrete DKP equations.
I first gather equations which remain independent after imposing the
determinant type consistency conditions.
To do it, I concentrate on equations ($\rf{lbte1}$).
When the determinant type consistency condition for these equations is
imposed, the rank of the coefficients matrix becomes 4.
Thus, out of 8 equations in ($\rf{dlbte}$), 4 equations remain independent.
I take such four equations as follows.
\be:
\and:
0:\si -z_{34}f_{134}g_{12}-z_{24}f_{124}g_{13}-z_{23}f_{123}g_{14}+z_{24}z_{
34}z_{23}f_{1}g_{1234}=0\nn
\and: 1:\si
z_{23}f_{234}g_{14}-z_{13}f_{134}g_{24}+z_{12}f_{124}g_{34}-z_{12}z_{23}z_{1
3}f_{4}g_{1234}=0\nn
\and:
2:\si -z_{24}f_{234}g_{13}+z_{14}f_{134}g_{23}-z_{12}f_{123}g_{34}+z_{14}z_{
24}z_{
12}f_{3}g_{1234}=0\nn
\and: 3:\si
z_{34}f_{234}g_{12}-z_{14}f_{124}g_{23}+z_{13}f_{123}g_{24}-z_{14}z_{34}z_{1
3}f_{2
}g_{1234}=0\nn
\lb{4dlb}
\ee:
However, these equations become dependent in ($\rf{lbte2}$) when the
determinant type consistency condition is imposed to it.
Namely, one equation out of 4 equations becomes a summation of other three
equations when f satisfies discrete DKP equations.
I choose latter three equations in ($\rf{4dlb}$) as such three equations.

Each of the three equations allow one to evaluate, from g on
$k_{1}+k_{2}+k_{3}+k_{4}=c$, g on $k_{1}+k_{2}+k_{3}+k_{4}=c+1$.
However, the conditions for g's obtained by each of the LBTE to coincide,
need to be imposed on initial value of g.
The conditions becomes as
\be:
\frac{z_{23}f_{234}g_{14}-z_{13}f_{134}g_{24}+z_{12}f_{124}g_{34}}{-z_{12}z_
{23}z_{14}f_{4}}\nn
=
\frac{-z_{24}f_{234}g_{13}+z_{14}f_{134}g_{23}-z_{12}f_{123}g_{34}}{z_{14}z_
{24}z_
{12}f_{3}}\nn
=
\frac{z_{34}f_{234}g_{12}-z_{14}f_{124}g_{23}+z_{13}f_{123}g_{24}}{-z_{14}z_
{34}z_
{13}f_{2}}\nn
\lb{dcon}
\ee:
Each of the equality represents the coincidence of $g_{1234}$ obtained by
three equation in ($\rf{4dlb}$).
These conditions require no more consistency condition.
Hence one can recursively construct the solution of g to LBTE, when the
obtained g also satisfies the same condition.
This is the case, when f satisfies the discrete DKP equations.
I show it for the first equality in equation ($\rf{dcon}$), as an example.
It can be shown by comparing following two procedures.
\be:
\and: g(B),g(A),g(C)\rightarrow g_{12}\nn
\and: g(A),g(D),g(E)\rightarrow g_{23}\sii
g_{12},g_{23},g_{34}\rightarrow g_{1234}\nn
\and: g(E),g(G),g(H)\rightarrow g_{34}\nn
\and: g(B),g(A),g(F)\rightarrow g_{14}\nn
\and: g(A),g(D),g(G)\rightarrow g_{24}\sii
g_{14},g_{24},g_{34}\rightarrow g_{1234}\nn
\and: g(C),g(E),g(H)\rightarrow g_{34}
\lb{dpr}
\ee:
Where,
\be:
A=(0,0,0,0) \si
B=(1,-1,0,0)\si
C=(0,-1,1,0) \nn
D=(-1,1,0,0)\si
E=(-1,0,1,0)\si
F=(0,-1,0,1)\nn
G=(-1,0,0,1)\si
H=(-1,-1,1,1)\si
\ee:
Here, I substitute 1 into c for simplicity, but it does not lose generality.
In these procedures , the value of $g(0,0,1,1)$ can be obtained by using
three equations in ($\rf{4dlb}$) independently.
This requires conditions, in the form of ($\rf{dcon}$), to be imposed on g
on the plane $k_{1}+k_{2}+k_{3}+k_{4}=-1$.
By using these conditions, one can omit $g(F)$ and $g(C)$ in ($\rf{dpr}$).
Equating the coefficients of remaining five g's on the plane
$k_{1}+k_{2}+k_{3}+k_{4}=-1$,
one can convert the condition into
\be:
g(G):
&&\!\!\!\frac{f(0,1,1,1)f(1,0,0,0)}{z_{14}^{2}z_{12}z_{13}z_{23}z_{14}z_{24}
z_{34}f(-1,0,1,1)f(0,0,0,1)}\nn
&\times&\!\!\! \Bigl[ (1-e^{-\partial_{1}-\partial_{2}})
\bigl( \frac{z_{14}z_{34}z_{13}f_{2}f_{134}-z_{14}z_{24}z_{12}f_{3}f_{124}+z
_{pq}z_{13}z_{23}f_{123}f_{4}}{f_{234}f_{1}}\bigr) \Bigl]\nn
&&\!\!\!=0
\ee:

\be:
g(E):
\and:
\frac{f(0,1,1,1)f(1,0,0,0)}{z_{14}z_{24}z_{34}z_{12}z_{13}^{2}z_{23}f(-1,0,0
,0)f(0,0,1,0)}\nn
\and: \times\Bigl[ (1-e^{-\partial_{1}-\partial_{2}})
\bigl( \frac{z_{13}z_{14}z_{34}f_{2}f_{134}+z_{12}z_{14}z_{24}f_{124}f_{3}-z
_{12}z_{13}z_{23}f_{4}f_{123}}{f_{1}f_{234}}\bigr)\Bigl]\nn
\and: =0
\ee:

\be:
g(A):
\and:
\frac{f(0,1,1,1)f(1,0,1,1)}{z_{14}z_{24}z_{12}^{2}z_{23}z_{13}f(0,0,0,1)f(0,
0,1,0)}\nn
\and: \Bigl[ (e^{-\partial_{1}}-e^{\partial_{2}})
\bigl(\frac{z_{14}z_{23}f_{14}f_{23}-z_{24}z_{13}f_{13}f_{24}+z_{34}z_{12}f_
{12}f_{34}}{ff_{1234}} \bigr)\Bigr]\nn
\and:
+\frac{f(0,1,1,1)f(1,0,0,0)f(-1,0,1,1)}{z_{14}^{2}z_{24}z_{12}z_{13}^{2}z_{2
3}f(-1,0,0,0)f(0,0,0,1)f(0,0,1,0)}\nn
\and: \Bigl[ (1-e^{-\partial_{1}-\partial_{2}})
\bigl(\frac{ -z_{14}z_{24}z_{12}f_{124}f_{3}+z_{12}z_{23}z_{13}f_{123}f_{4}+
z_{14}z_{34}z_{13}f_{134}f_{2}}{f_{1}f_{234}}\bigr)\Bigr]\nn
\and: =0
\ee:
All these equations are satisfied if f satisfies two discrete DKP equations.
The equality of equation 2 and 3 or 1 and 3 in ($\rf{dcon}$), can be shown
by using similar approach.
Thus, it is shown that, whenever f satisfies the discrete DKP equations,
there
is at least one solution of g to LBTE.
\\

\appendix
\section{Appendix: discrete formula for the integrable hierarchy}

Integral identity known in the name of bilinear identity
\cite{KP}\cite{BKP},
leads all the equations satisfied by the $\tau$-function of the integrable
hierarchy.
The discrete formulae that appear in this paper is also obtained from this
identity.
In this appendix,
I briefly explain the way to obtain the discrete formulae from the bilinear
identities,
for completeness.

\subsection{Bilinear identity for the KP hierarchy and the Modified KP hierarchy }
Bilinear identity for the KP hierarchy is as follows \cite{KP}.
\be:
\oint \frac{dz}{2 \pi i}
e^{\xi(x,z)-\xi(x',z)}\tau(x-\epsilon(z^{-1}))\tau(x'+\epsilon(z^{-1}))=0
\lb{kpi}
\ee:
Here, $x$ and $x'$ are infinite dimensional vectors emboding infinite number
of variables in the KP hierarchy.
\be:
x=(x_{1},x_{2},x_{3},.....)
\lb{cv}
\ee:
These variables are continuous variables in the sense that they become
 variables for differential equations contained in the KP hierarchy.
$\epsilon(z)$ is a vector represented as
\be:
\epsilon(z)=(z,\frac{1}{2}z^{2},....)
\ee:

$\xi$ is the function of x and z represented as
\be:
\xi(x,z)=\sum_{n=1}^{\infty}\frac{1}{n}x_{n}z^{n}
\ee:
By choosing $x-x'$ properly,
one can obtain all the equations satisfied by tau-function for the KP
hierarchy.

Infinite number of discrete variables are defined through Miwa
transformation
\cite{KP} from continuous variables ($\rf{cv}$).
\be:
\frac{\partial}{\partial k_{i}}=\sum_{n=1}^{\infty}\frac{1}{n}
\frac{\partial}{\partial x_{n}}z_{i}^{n}
\ee:
Here, $z_{i}$'s are arbitrary complex constant and discrete variable exist
for
every different $z_{i}$'s.

Difference equations contained in the KP hierarchy depend on these
variables.

The differnce equation itself is obtained by expanding bilinear identity,
after substituting appropriate summation of
$\epsilon(z_{1}),\epsilon(z_{2}),....$ to $x-x'$.
Dimension of the equations increases with the number of the terms contained
in the summation .
Most simple nontrivial equation in it is obtained when one substitute
$\epsilon(z_{1})+\epsilon(z_{2})+\epsilon(z_{3})$ to $x-x'$.
After some calculation one obtains following equation, in this case.
\be:
z_{1}z_{23}\tau_{k_{1}} \tau_{k_{2}k_{3}}
+z_{2}z_{31}\tau_{k_{2}} \tau_{k_{3}k_{1}}
+z_{3}z_{12}\tau_{k_{3}} \tau_{k_{1}k_{2}}
=0
\lb{kp}
\ee:
Hence, the lowest dimensional discrete equation contained in the KP
hierarchy is
a three dimensional equation.
Another three dimensional equation is obtained, when one substitute
$\epsilon(z_{1})+\epsilon(z_{2})+\epsilon(z_{3})-\epsilon(z_{4})$ to $x-x'$.
That is the discrete KP equation employed in this paper.
Replacing $z_{i4}$ by $z_{i}(i=1,2,3)$, after suitable variable
transformation, one can convince oneself that the discrete KP equation is
same
as equation ($\rf{kp}$).

Bilinear identity for the modified KP hierarchy can be obtained as a
modification of the KP hierarchy's one \cite{KP}.
It takes the following form.
\be:
Res_{z=0}dz z
e^{\xi(x,z)-\xi(x',z)}\tau'(x-\epsilon(z))\tau(x'+\epsilon(z))=0
\ee:
By substituting summations of arbitary three of
$\epsilon(z_{1}),\epsilon(z_{2}),\epsilon(z_{3}),\epsilon(z_{4})$ to $x-x'$
this identity leads to 4 equations same as those in LBTE for the discrete KP
equation.

\subsection{Bilinear identity for the BKP hierarchy and the Modified BKP hierarchy}
Tau-function for the BKP hierarchy satisfies the following bilinear identity
\cite{KP}.
\be:
Res_{z=0}dz\frac{1}{z}
e^{\tilde{\xi}(x,z)-\tilde{\xi}(x',z)}\tau(x-2\tilde{\epsilon}(z^{-1}))\tau(
x+2\tilde{\epsilon}(z^{-1}))
=\tau(x)\tau(x')
\ee:

Here, $x,x',\tilde{\epsilon}$ and $\tilde{\xi}$ are same as those for the KP hierarchy except
the
absence of variables of even number indices. i.e.
\be:
\and: x=(x_{1},x_{3},x_{5},....)\nn
\and: \tilde{\epsilon}(z)=(z,\frac{1}{3}z^{3},....)\nn
\and: \tilde{\xi}(x,z)=\sum_{n=1}^{\infty}\frac{1}{2n-1}x_{2n-1}z^{2n-1}
\ee:
By expanding this bilinear formula,
one can obtain all the equations contained in the BKP hierarchy.

Dependent variables for discrete equations which belong to the BKP hierarchy are
difined through following variable transformation.
\be:
\frac{\partial}{\partial k_{i}}=2\sum_{n=1}^{\infty}\frac{1}{2n-1}
\frac{\partial}{\partial x_{2n-1}}z_{i}^{2n-1}
\ee:
The discrete equations themselves are obtained by substituting the summation
of
$\tilde{\epsilon}(z_{1}),\tilde{\epsilon}(z_{2}),\tilde{\epsilon}(z_{3})...$
to
$x-x'$.
Most simple nontrivial equation is obtained when one substitutes
$\tilde{\epsilon}(z_{1})+\tilde{\epsilon}(z_{2})+\tilde{\epsilon}(z_{3})$ to
$x-x'$.
The equation is the discrete BKP equation.

Bilinear identities for the Modified BKP hierarchy can be obtained as a
modification of those for the BKP hierarchy and takes the following form
\cite{BKP}.
\be:
Res_{z=0}dz \frac{1}{z}
e^{\tilde{\xi}(x,z)-\tilde{\xi}(x',z)}\tau'(x-2\tilde{\epsilon}(z^{-1}))
\tau(x+2\tilde{\epsilon}(z^{-1}))
=2\tau(x)\tau'(x')-\tau'(x)\tau(x)
\ee:

By substituting summation of two of
$\tilde{\epsilon}(z_{1}),\tilde{\epsilon}(z_{2}),\tilde{\epsilon}(z_{3})$ to
$x-x'$, one can obtain LBTE for discrete BKP equation used in this paper.

\subsection{Bilinear identity for the fermionic BKP hierarchy}

Bilinear identity for the fermionic BKP hierarchy is\cite{Kac}
\be:
\frac{1}{2}((-1)^{n+m}-1)\tau_{n}(x)\tau_{m}(x')\nn
+Res dz
z^{n-m-2}e^{\xi(x,z)-\xi(x',z)}\tau_{n-1}(x-\epsilon(z))
\tau_{m+1}(x'+\epsilon(z))\nn
+Res dz
z^{m-n-2}e^{\xi(x,z)-\xi(x',z)}\tau_{n+1}(x+\epsilon(z))
\tau_{m-1}(x'-\epsilon(z))
=0
\ee:
where $x,x',\epsilon(z)$ and $\xi(x,z)$ are same as those of the KP
hierarchy,
and n represents the charge in the free fermion description \cite{KP}
\cite{BKP}.
Variables for discrete equaiton are also same as those of the KP hierarchy.
All the equations which belong to the BKP hierarchy can be obtained by
chosing
$x-x'$ and $n-m$ appropriately.
Most simple nontrivial discrete equaiton is obtained when
$x-x'=\epsilon(z_{1})+\epsilon(z_{2})$ and $n-m=1$ or $3$, and is
\be:
z_{2}\tau_{n+1}(p+1,q)\tau_{n+2}(p,q+1)+z_{1}\tau_{n+1}(p,q+1)\tau_{n+2}(p+1
,q)\nn
+z_{12}\tau_{n+1}(p,q)\tau_{n+2}(p+1,q+1)+z_{12}z_{1}z_{2}\tau_{n}(p,q)\tau_{
n+3}(p+1,q+1)=0
\ee:
This equation is three dimensional equation when one considers n as one
variable.
To make the three variables equal footing,
I transform these variables as follows.
\be:
p=k_{1}, q=k_{2}, n=p+q
\ee:
In these variables, the equation is expressed as
\be:
z_{2}f_{1}f_{23}
+z_{1}f_{2}f_{13}
+z_{12}f_{3}f_{12}
+z_{12}z_{2}z_{1}f f_{123}=0
\ee:
This equation is three dimensional discrete equaiton.
However another three dimensional equaiton can be obtained when
$x-x'=\epsilon(z_{1})+\epsilon(z_{2})-\epsilon(z_{3})$ and $n-m=1$.
The equation is discrete fermionic BKP equation employed in this paper.

Bilinear identity for the Modified fermionic BKP hierarchy is
\be:
Res dz
z^{n-m-2}e^{\xi(x,z)-\xi(x',z)}\tau_{n-1}'(x-\epsilon(z))
\tau_{m+1}(x'+\epsilon(z))\nn
+Res dz
z^{m-n-2}e^{\xi(x,z)-\xi(x',z)}\tau_{n+1}'(x+\epsilon(z))
\tau_{m-1}(x'-\epsilon(z))\nn
=-\frac{1}{2}((-1)^{n+m}-1)\tau_{n}'(x)\tau_{m}(x')+\tau_{n}(x)\tau'_{m}(x')
\ee:
By substituting arbitrary sums of two of
$\epsilon(z_{1}),\epsilon(z_{2}),\epsilon(z_{3})$ into $x-x'$, one can
obtain three equations in LBTE for the discrete fermionic BKP equaiton.

\subsection{Bilinear identity for the DKP hierarchy }
Bilinear identity for the DKP hierarchy is as follows \cite{Kac}.
\be:
\and: Res_{z=0}dz z^{n-m-2}e^{\xi(x,z)-\xi(x',z)}
\tau_{n-1}(x-\epsilon(z^{-1}))\tau_{m+1}(x'+\epsilon(z^{-1}))\nn
\and: +Res_{z=0}dz z^{m-n-2}e^{-\xi(x,z)+\xi(x',z)}
\tau_{n+1}(x+\epsilon(z^{-1}))\tau_{m-1}(x'-\epsilon(z^{-1}))
=0
\ee:
Here, n represents the charge in the free fermion description.
$x,x',\epsilon$ and $\xi$ are same as the KP hierarchy.
Dependent variables for discrete equations which belong to the DKP hierarchy
are
also same as the KP hierarchy.
All the equations which belong to the DKP hierarchy can be obtained by
expanding
this identity, when one choose $n-m$ and $x-x'$ approariately.
Lowest dimensional discrete equations in it are obtained when
$x-x'=\epsilon(z_{1})+\epsilon(z_{2})+\epsilon(z_{3})$ and $n-m=2,4$ or
$x-x'=\epsilon(z_{1})+\epsilon(z_{2})-\epsilon(z_{3})$ and $n-m=0,2$.
However, two of these equations overlap with other two equations.
As a consequence, the following two equations turn out to be the lowest
dimensional discrete DKP equations.
\be:
z_{p}z_{qr}f_{n}(p,q+1,r+1)f_{n}(p+1,q,r)
+z_{q}z_{rp}f_{n}(p+1,q,r+1)f_{n}(p,q+1,r)\nn
+z_{r}z_{pq}f_{n}(p+1,q+1,r)f_{n}(p,q,r+1)
-z_{p}z_{q}z_{r}z_{pq}z_{qr}z_{rp}f_{n+2}(p+1,q+1,r+1)f_{n-2}(p,q,r)=0\nn
z_{q}z_{r}z_{qr}f_{n-1}(p,q+1,r+1)f_{n-3}(p+1,q,r)
+z_{r}z_{p}z_{rp}f_{n-1}(p+1,q,r+1)f_{n-3}(p,q+1,r)\nn
+z_{p}z_{q}z_{pq}f_{n-1}(p+1,q+1,r)f_{n-3}(p,q,r+1)
+z_{pq}z_{qr}z_{rp}f_{n-1}(p+1,q+1,r+1)f_{n-3}(p,q,r)=0
\lb{ddkp}
\ee:
These equations are 4 dimensional equations, considering n as one variable.
To make the four variables equal footing, I transform these four variables
as follows.
\be:
p=k_{1},q=k_{2},r=k_{3},n=k_{1}+k_{2}+k_{3}+k_{4}
\ee:
In these four variables, the equations take the following forms.
\be:
z_{1}z_{23}f_{23}f_{14}
-z_{13}z_{2}f_{13}f_{24}
+z_{12}z_{3}f_{12}f_{34}
-z_{12}z_{13}z_{1}z_{23}z_{2}z_{3}f_{1234}f=0\nn
z_{23}z_{2}z_{3}f_{23}f_{1}
-z_{13}z_{1}z_{2}f_{134}f_{2}
+z_{12}z_{1}z_{2}f_{124}f_{3}
-z_{12}z_{13}z_{23}f_{123}f_{4}=0
\ee:
However, one can obtain other four dimensional equations for the DKP
hierarchy
as in the previous case.
One of which is obtained when
$x-x'=\epsilon(z_{1})+\epsilon(z_{2})+\epsilon(z_{3})-\epsilon(z_{4})$ and
$n-m=2$, and is the first equation in the dicrete DKP equations employed in
this
paper.
Another one is obtained when
$x-x'=\epsilon(z_{1})+\epsilon(z_{2})+\epsilon(z_{3})+\epsilon(z_{4})$ and
$n-m=4$ and is the second equation in the discrete DKP equation.

Modification of bilinear identity for the discrete DKP equation leads
\cite{Kac}
\be:
\and: Res_{z=0}dz z^{n-m-2}e^{\xi(x,z)-\xi(x',z)}
\tau'_{n-1}(x-\epsilon(z^{-1}))\tau_{m+1}(x'+\epsilon(z^{-1}))\nn
\and: +Res_{z=0}dz z^{m-n-2}e^{-\xi(x,z)+\xi(x',z)}
\tau'_{n+1}(x+\epsilon(z^{-1}))\tau_{m-1}(x'-\epsilon(z^{-1}))\nn
\and: =\tau_{n}(x)\tau'_{m}(x')
\ee:
This identity is bilinear identity for the Modified DKP equation.
By taking arbitrary summations of three of
$\epsilon(z_{1}),\epsilon(z_{2}),\epsilon(z_{3}),\epsilon(z_{4})$, one can
obtain 8 equations in LBTE for the discrete DKP equations.
\\
{\Large{\bf \noindent Acknowledgement}}\\
I wish to thank to Professor Igor Korepanov and Professor S. Saito for fruitful discussion


\end{document}